\documentclass[twocolumn,showpacs,preprintnumbers,amsmath,amssymb]{revtex4}
\usepackage{graphicx}
\usepackage{dcolumn}
\usepackage{latexsym}
\usepackage{bm}
\begin{document} 
\title{Variational method for lattice
spectroscopy with ghosts}

\author{Tommy Burch$^a$}
\author{Christof Gattringer$^b$} 
\author{Leonid Ya.\ Glozman$^b$}\
\author{Christian Hagen$^a$}
\author{C. B. Lang$^b$}
\affiliation{\vspace{1mm}{\rm(Bern-Graz-Regensburg (BGR)
collaboration)}\vspace{2mm}\\
$^a$Institut f\"ur Theoretische Physik, Universit\"at
Regensburg \\
D-93040 Regensburg, Germany\vspace{2mm}}
\affiliation{$^b$Institut f\"ur Physik, FB Theoretische Physik
Universit\"at Graz \\
A-8010 Graz, Austria\vspace{2mm}}
\date{November 23, 2005}

\begin{abstract}
We discuss the variational method used in lattice spectroscopy calculations. 
In particular we address the role of ghost contributions which appear 
in quenched or partially quenched simulations 
and have a non-standard euclidean time dependence. 
We show that the ghosts can be separated from the physical states. Our 
result is illustrated with numerical data for the scalar meson.
\end{abstract}

\maketitle

\section{Quenched correlators and their spectral representation}

Spectroscopy of excited states is still a quite challenging problem for 
lattice QCD. The reason is that euclidean two-point functions 
have a sum of infinitely many exponentials in their spectral representation
and it is a non-trivial task to extract the subleading terms 
corresponding to the excited states. 
For quenched or partially quenched calculations the situation is complicated further by 
additional unphysical contributions: When ignoring the fermion determinant, 
the $\eta^\prime$ also becomes a Goldstone boson, 
and this massless flavor singlet 
state can couple in various processes through hairpin diagrams. 
The corresponding effects were studied in \cite{BaDuEi01} for the 
quenched scalar propagator. It was demonstrated that the $\eta^\prime$ 
leads to additional contributions
to the scalar propagator which have a negative sign and a non-standard
$t$-dependence. Such terms, often referred to as ``ghosts'', compete
with the exponential decay coming from the scalar, and make the extraction of 
the scalar mass cumbersome. For the nucleon system the effects of 
ghost states and the problems they cause in the extraction of the Roper
resonance were discussed in \cite{MaChDo05}.

In principle, in a quenched calculation,
correlators for all quantum numbers may be infested by ghosts, 
possibly coming from bound or scattering states
with $\eta^\prime$, or states involving even several $\eta^\prime$. 
Thus, one has to find tools for reliably removing 
these contributions, in
particular, if one is interested in excited states, where the influence of
the ghost states is even more devastating. 

A powerful tool, developed for the analysis of excited states, 
is the variational method \cite{Mi85,LuWo90}. 
The main idea is to work with several linearly independent 
interpolators $O_i, \, i = 1, 2, \, \ldots \, r$, 
all with the quantum numbers of the desired
state, and to compute the cross-correlations
\begin{equation}
C(t)_{ij} \; = \; \langle \, O_i(t) \, \overline{O}_j(0) \, \rangle \;
\; , \; \; \; i,j \; = 1, 2, \, \ldots \, r .
\label{corrmatdef}
\end{equation}
In full euclidean lattice QCD the correlators have the spectral decomposition
\begin{equation}
C(t)_{ij} \; = \; \sum_{n=1}^\infty v_i^{(n)} v_j^{(n)^*}
 \, e^{-t \, V_n}  \; ,
\label{corrhilbert}
\end{equation}
where the coefficients $v_i^{(n)}$ are given by 
\begin{equation} 
v_i^{(n)} \; = \; \langle \, 0 \, | \, O_i \, | \, n \, \rangle \; .
\label{coeffs}
\end{equation}
In Eq.\ (\ref{corrhilbert}) $V_n$ denotes the energy of the state
$|n\rangle$.

In the presence of ghosts
we must augment the decomposition (\ref{corrhilbert}) by
additional terms which take into account the possible negative sign
and the different $t$-dependence of the ghost contributions. 
Since, in principle, there are infinitely
many ghost states, their contribution also gives rise to an infinite tower 
of states  
\begin{equation}
\sum_{n=1}^\infty w_i^{(n)} w_j^{(n)^*}
 \, f^{(n)}(t) \, e^{-t \, W_n}  \; .
\label{corrhilbert2}
\end{equation}
Here $W_n$ is the energy of the $n$-th ghost state and
the function $f^{(n)}(t)$ describes the non-exponential part of its euclidean time dependence, 
taking into account also a possible negative sign. 
We stress that for our analysis we do not have to know the precise form of 
$f^{(n)}(t)$. Certainly it is not simple to present a form which
encompasses all possibilities that might occur.
However, to give an example, we remark that the linear behavior
\protect{$f(t)=-(a+bt)$} was suggested for the system considered in 
\cite{MaChDo05}. 

In order to combine the contribution (\ref{corrhilbert2}) from the ghosts
and the original decomposition (\ref{corrhilbert}), we represent our 
correlation matrix as
\begin{equation}
C(t)_{ij} \; = \; \sum_{n=1}^\infty a_i^{(n)} a_j^{(n)^*}
 \, H^{(n)}(t) \, e^{-t \, E_n}  \; ,
\label{corrdeco}
\end{equation} 
where
\begin{eqnarray}
H^{(n)}(t)  &=&  1 \; , \; E_n = V_{n_P}  
\; \; \; \mbox{for proper states} \; ,
\nonumber 
\\
H^{(n)}(t)  &=&  f^{(n_G)}(t) \; , \;  E_n =  W_{n_G} 
\; \; \; \mbox{for ghosts} \; .
\end{eqnarray}
Here we have ordered the energy levels such that (assuming non-degenerate 
values for the $E^{(n)}$)
\begin{equation}
0 \; < \; E_1 \; < \; E_2 \, < \;  E_3 \; \ldots \; ,
\end{equation}
implying that the ghost state energies are interleaved with the energies 
of proper states. A typical situation would be
\begin{equation}
0 \; < \; V_1 \; < \; W_1 \, < \;  V_2 \; \ldots \; ,
\end{equation}
where the energy $W_1 \equiv E_2$ of the first ghost state comes after the
ground state energy $V_1 \equiv E_1$, but is below the first excited energy
level $V_2 \equiv E_3$. In such a case the ghost contribution overlays 
the exponential decay from $V_2$ and makes the extraction of this number 
from a single correlator quite difficult.

\section{Variational method with ghost contributions}

Having generalized the formula for the spectral representation 
from the original
form (\ref{corrhilbert}) to the ansatz (\ref{corrdeco}), which is capable of
describing also the ghost contributions, we can now proceed as in 
\cite{LuWo90}. In particular, we consider the generalized eigenvalue problem
\begin{equation}
C(t) \, \vec{\psi}^{(k)} \; \; = \; \; \lambda(t)^{(k)} \, C(t_0) \, 
\vec{\psi}^{(k)} \; .
\label{generalized}
\end{equation}
On the right-hand side we use the correlation matrix at some fixed
$t_0 < t$ to normalize the eigenvalue problem. 
We now show that the eigenvalues $\lambda(t)$ of the generalized
eigenvalue problem are given by
\begin{eqnarray}
&&\!\!\!\!\!\! \lambda^{(k)}(t)  =  e^{- (t - t_0)E_k}  
\frac{H^{(k)}(t)}{H^{(k)}(t_0)} 
\Big[1 + {\cal O}(e^{-(t - t_0) \Delta_k}) \Big] \!\!\!\!\!\!
\label{eigenvaluedecay}
\\
&&\!\!\!\!\!\!  =   
\left\{\! \begin{array}{l} 
e^{- (t - t_0) V_k} 
\Big[ 1 + {\cal O}(e^{-(t-t_0) \Delta_k}) \Big] \; \;
 \mbox{for proper states} , \\
\\
e^{- (t - t_0) W_k} \frac{f^{(k)}(t)}{f^{(k)}(t_0)}  
\Big[ 1 + {\cal O}(e^{-(t-t_0) \, \Delta_k}) \Big] \;\;
\mbox{for ghosts} ,
\end{array} \right. 
\nonumber
\end{eqnarray}
where $E_k$ is the energy of the $k$-th state (proper state or
ghost) and $\Delta_k = E_{r+1} - E_k$
is the difference to the mass of the $r\!+\!1$-st state, 
$r$ being the number of interpolators used for the correlators.

The proof proceeds by considering the generalized eigenvalue problem 
\begin{equation}
\widetilde{C}(t) \, \vec{\psi}^{(k)} \; \; = \; \; \lambda(t)^{(k)} \, 
\widetilde{C}(t_0) \, 
\vec{\psi}^{(k)} \; ,
\label{generalizedred}
\end{equation}
for the hermitian $r\times r$ matrix $\widetilde{C}(t)$,
obtained by truncating the spectral sum after the $r$-th term,
\begin{equation}
\widetilde{C}(t)_{ij} = \sum_{n=1}^r a_i^{(n)} a_j^{(n)^*}
 \, H^{(n)}(t) \, e^{-t \, E_n}  
\; , \; \; \; i,j \; = 1, 2, \, \ldots \, r .
\label{corrdecored}
\end{equation} 
We assume that the coefficients $a_i^{(n)}, \, i,n = 1,2, \,
\ldots \, r$, form a matrix of full rank. For the proper states the $a_i^{(n)}$
are given by (\ref{coeffs}). For linearly independent interpolators $O_i$,
that couple to the first $r$ states,
full rank follows from the fact that the states $| n \rangle$ are orthonormal.
For the quenched approximation one expects that the ghost states are
linearly independent of the proper states such that again for linearly
independent $O_i$ the matrix $a_i^{(n)}$ has full rank. 

Inserting (\ref{corrdecored}) in (\ref{generalizedred}) one finds
that for all $i$
\begin{equation}
\sum_{n=1}^r\!a_i^{(n)} \rho^{(n,k)}
\Big[ H^{(n)}(t) e^{-t E_n} - \lambda(t)^{(k)} H^{(n)}(t_0) e^{-t_0 \, E_n} 
\Big]\! = 0,
\label{deriv1}
\end{equation}
with the coefficients $\rho^{(n,k)}$ given by
\begin{equation}
\rho^{(n,k)} \; = \; \sum_{j=1}^r a_j^{(n)^*} \, \psi_j^{(k)} \; .
\label{deriv2}
\end{equation}
From the full rank of $a_i^{(n)}$ it follows that the vectors $a_i^{(n)}, \,
n = 1,2, \, \ldots \, r$, are linearly independent, implying
\begin{equation}
\rho^{(n,k)}
\, \Big[ H^{(n)}(t) e^{-t E_n} - \lambda(t)^{(k)} H^{(n)}(t_0) e^{-t_0 E_n} 
\Big] = 0 \;  \; \forall \; n .
\label{deriv3}
\end{equation}
Using the linear independence of the 
eigenvectors $\vec{\psi}^{(k)}$ and the full rank of $a_j^{(n)}$, it follows
that for each $n$ there exists a $k$, such that $\rho^{(n,k)}$ 
is non-vanishing. Consequently, Eq.\ (\ref{deriv3}) implies that the
eigenvalues of the reduced problem (\ref{generalizedred}) are 
given exactly by the leading term in (\ref{eigenvaluedecay}). 
The full matrix $C(t)$ is
obtained from $\widetilde{C}(t)$ by adding terms of ${\cal O}(\exp(-t \;
E_{r+1}))$ which can be treated perturbatively \cite{LuWo90}, giving rise to 
the correction in (\ref{eigenvaluedecay}).

Before we come to discussing an application of the generalized variational 
method, let us briefly address once more the implications of the result 
(\ref{eigenvaluedecay}): In the leading term each eigenvalue couples to only 
one state (ghost or proper state). Thus, ghosts can be cleanly
disentangled from the proper states, up to the 
correction term ${\cal O}(\exp(-t \; \Delta_k))$. We stress that the form 
of the euclidean time dependence for the ghosts, $f^{(k)}(t)$, needs not be 
known or modelled for the application of the method. In the example we
discuss below, the ghost state is easily identified by its non-standard 
$t$-dependence.

\begin{figure*}[t]
\begin{center}
\hspace*{-6mm}
\includegraphics*[width=165mm]{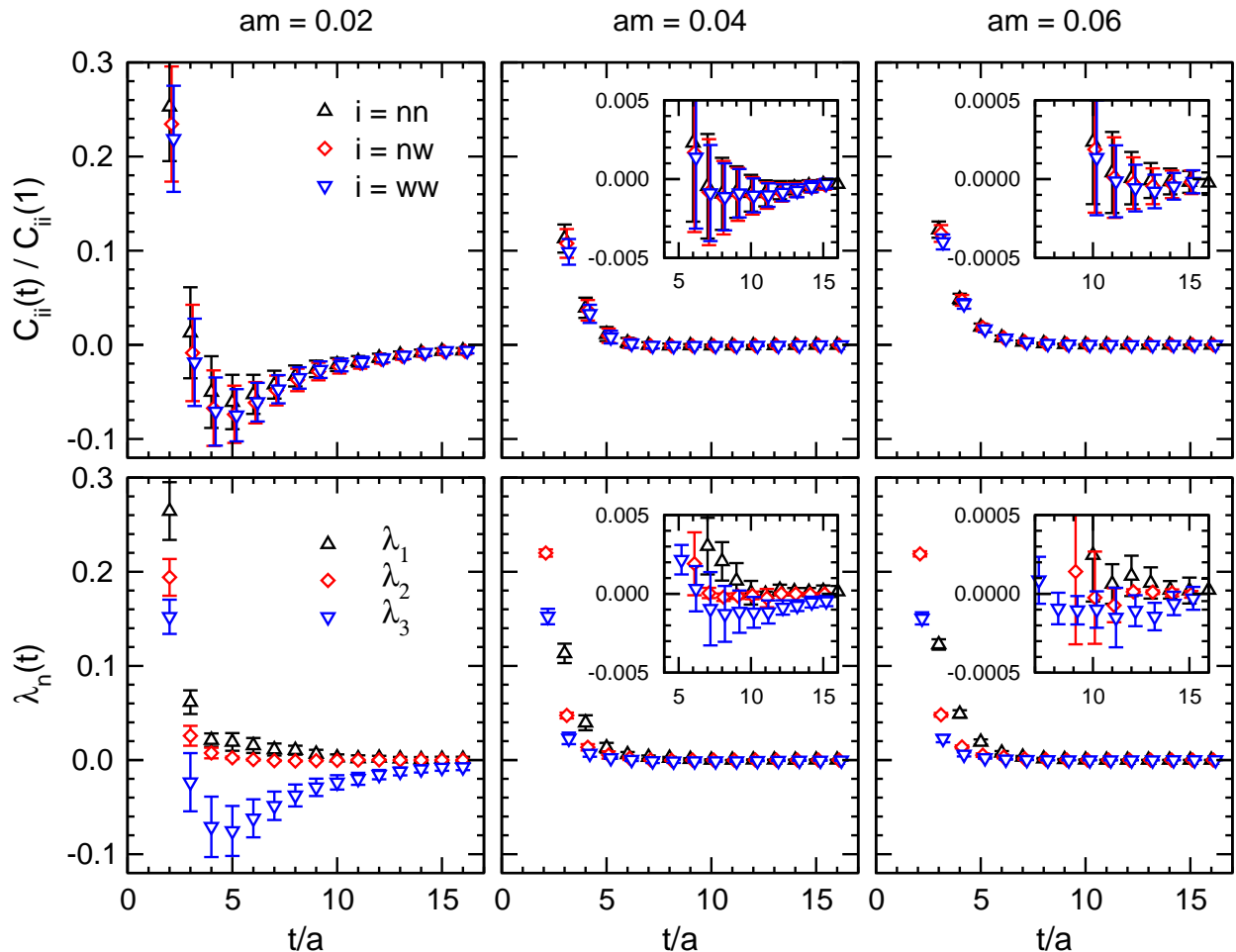} 
\end{center}
\caption{Euclidean time dependence for the diagonal
entries of the correlation matrix
(top row) and for the eigenvalues of the 
generalized eigenvalue problem
(bottom row). The quark masses are
$am = 0.02, 0.04$ and 0.06 (left to right).} 
\label{fig1}
\end{figure*}

We remark that additional information for the identification of the ghost states
comes from the standard eigenvalue problem, i.e., when the normalization with 
$C(t_0)$ is omitted on the right-hand side of
(\ref{generalized}). Following the arguments given in the
appendix of \cite{LuWo90}, one finds that for the standard eigenvalue problem
the eigenvalues are given by
\begin{equation}
\lambda^{(k)}(t)  \propto  e^{- t E_k}  
H^{(k)}(t) 
\Big[1 + {\cal O}(e^{-t \delta_k}) \Big] \; .
\label{standardev}
\end{equation}
Since here the factor $H^{(k)}(t)$ is not normalized by $H^{(k)}(t_0)$, 
the sign information is not lost and an overall sign indicates 
a ghost. However, this information is obtained at the cost of a much larger 
correction term, since in (\ref{standardev}) $\delta_k$ is the distance
to the nearest energy level which is usually much smaller than 
$\Delta_k$ appearing in (\ref{eigenvaluedecay}). 

\section{Illustration of the result for the scalar meson propagator}

We illustrate our result using the scalar meson 
as an example. In particular, we work with the three interpolators
\begin{equation}
\overline{u}_n \, d_n \; , \; 
\overline{u}_n \, d_w \; , \; 
\overline{u}_w \, d_w \; .
\label{scalarbasis}
\end{equation}
The subscript $n$ denotes a narrow quark source (or sink),
while $w$ is used for wide. The narrow and wide quark sources (sinks) 
are constructed via
different amount of Jacobi smearing \cite{jacobi}. 
Such interpolators with mixed width for the quark sources have been 
utilized in lattice spectroscopy of excited mesons and baryons
and details of the source preparation can be found in \cite{bgrexcite}.

We remark, that the interpolators used here are not meant to 
resolve the issues currently discussed for the scalar meson (see e.g.\
\cite{BaDuEi01,scalarmeson}). They only serve to illustrate our result for the system
where the importance of ghost contributions 
was first discussed \cite{BaDuEi01}. A detailed analysis of the spectroscopy
results for the interpolators (\ref{scalarbasis}) will be presented elsewhere.

Using the three interpolators (\ref{scalarbasis}) we work with a $3 \times 3$ 
correlation matrix $C(t)$. We evaluate it on 100 quenched gauge 
configurations on a $20^3 \times 32$ lattice using the chirally improved
lattice Dirac operator \cite{chirimp}. The gauge action is the L\"uscher-Weisz
action at $\beta = 8.15$, corresponding to a lattice spacing of $a = 0.119$ 
fm determined from the Sommer parameter \cite{scale}. 
We present results at three
different quark masses $am = 0.02, 0.04, 0.06$, giving rise to pion masses 
of $m_\pi \sim 400, 550$ and 670 MeV \cite{bgr}. 
All errors given are statistical errors 
determined with single elimination jackknife.

In the top row of plots in Fig.\ \ref{fig1} 
we show the euclidean 
time dependence for the diagonal entries of the correlation matrix. We remark
that the off-diagonal elements show qualitatively the same behavior, but we
omit them in the plot to avoid overcrowding it. For the smallest quark mass 
the entries of the correlation matrix become negative at $t/a = 3$ or 4,
showing the strong influence of the ghost contribution. For $am = 0.04$ the
effect is still visible, but the correlators become negative only at values
$t/a = 6$ or 7. At the largest quark mass the correlators do not become 
significantly negative, i.e., within error bars are compatible with zero  
for $t/a > 10$. This indicates that the role of the ghost state is 
much weaker at larger quark masses. This is as expected from the fact, that
the mass of the would-be Goldstone particles, and thus of the ghost states,
increases very fast with the quark mass.

The bottom row plots show the time dependence
for the three eigenvalues of the generalized eigenvalue problem. They behave 
as expected from formula (\ref{eigenvaluedecay}). Only one of them,
$\lambda^{(3)}$, shows the behavior characteristic of
ghosts. The other two eigenvalues $\lambda^{(1)}$ and $\lambda^{(2)}$ 
remain positive and are compatible with a single exponential decay. 
The sign change of $\lambda^{(3)}$ is most prominent at $am = 0.02$ and 
the overall time dependence is similar to the behavior of the corresponding 
diagonal elements of the correlation matrix in the top left plot.
For $am = 0.04$ the ghost-like behavior of $\lambda^{(3)}$ is still visible,
while for $am = 0.06$ it is no longer significant.

Comparing the top and bottom row plots 
of Fig.\ 1 illustrates the strength 
of the method: For the two masses $am = 0.02$ and 0.04,
where the single correlators show a clear ghost contribution, the ghost 
appears in only one of the eigenvalues, while the other two are consistent
with a dominant single exponential behavior. Our result shows that with the
variational method ghost and proper states each dominate 
individual eigenvalues and the masses of the physical states can be fit without
having to know or model the form of the ghost contribution. 

\newpage
\noindent
{\bf Acknowledgements:} 
The calculations were done on the Hitachi SR8000
at the Leibniz Rechenzentrum in Munich and we thank the LRZ staff for
training and support. L.~Y.~G is supported by ``Fonds zur F\"orderung
der Wissenschaftlichen Forschung in \"Osterreich'', FWF, project P16823-N08.
This work is supported by DFG and BMBF. We thank D.\ Aykroyd and H.\  Ramis for sharing their
experience with removing ghosts.

\end{document}